\documentclass[10pt, twocolumn]{article}
\usepackage[utf8]{inputenc}
\usepackage[OT4]{fontenc}
\usepackage{amsmath}
\usepackage{amssymb}
\usepackage{graphicx}
\usepackage{multirow}
\usepackage{array}
\usepackage{subfigure}

\title{Classes of states of discrete systems}

\author{Malgorzata J. Krawczyk\\
Faculty of Physics and Applied Computer Science,\\
AGH University of Science and Technology,\\
al. Mickiewicza 30, 30-059 Krakow, Poland\\
Malgorzata.Krawczyk@agh.edu.pl}

\date{\today}

\begin{document}

\maketitle
\abstract{We present a method which allows for a reduction of a size of a simulated system. The method can be applied to any system where one can define a finite set of possible states of the system and an elementary process which transforms one state of the system to another. The method is based on the system symmetry; we get classes of states, which can be used instead of states. A detailed procedure is presented for undirected/directed and/or unweighted/weighted graphs. Several applications of the method are shown for different types of systems. In particular, new results are presented for the random Boolean networks.}

\section{Introduction}
The question we address here is if it is possible to indicate a representation of the system, we are interested in, which allows for a reduction of the system size with no loss of information. Our solution of this problem is based on the system symmetry which allows us to divide the state space of the system into sets of similar states. If it is the case, the system can then be represented by these sets. The proposed method is applicable for systems described by discrete states. Here, any two states of the system are connected if there is an elementary process which transforms one state into another. Such a description allows for the construction of a network whose nodes are identified with particular states of the system, and edges represent possible transitions between the states. The graph obtained in this way reflects the considered structure, i.e. the pattern of connections between the states of the analysed system. The graph is equivalent to a Kripke structure \cite{kri}. In the simplest case, the graph is undirected and unweighted. However, more than often it is not the case, and not all transitions are equally probable. This leads to an application of weighted graphs. It is also possible that weights of transitions between two states in both directions are not the same; in a limit case the transition in one direction may not be possible.\\
The simplest observation one can make on the similarity of states concerns the number of nearest neighbours of each node of the obtained graph. However, it is possible that the properties of a particular node depend also on its connections with further neighbours. We can take this option indirectly into account by an analysis of the neighbourhood of the nearest neighbours of each node. Such an analysis allows us to indicate nodes for which the same structure of connections is observed. We say, that those nodes belong to the same class of nodes. To summarize, two nodes belong to the same class if they have the same number of nearest neighbours which belong to the same classes. The existence of classes of nodes is an indication of the symmetry of the state space.\\
The advantage of the class structure is that the obtained classes can be used instead of states to represent the analysed system. If the system walks randomly in the state space, after long time all states which belong to a given class have the same probability. Thus, one can construct a new graph whose nodes represent classes, and edges represent connections between states which form a given class. The transformation from the initial to the new network of states preserves the stationary probability distribution of the system; hence the calculations of the probability distribution can be carried out for a much smaller system. The presented method is general, and it can be applied to any system which can be described as a graph of states connected by elementary processes.\\

Below we present a summary of our method, with an indication of its crucial elements. We present results of the application of the method to the state space formed by the set of all possible states of Boolean networks of a given size. We also show some other examples of the application the method. In each case, the method results in a significant reduction of the system size. The paper is organised as follows: In the next section the class identification method is presented. The following two section are devoted to the presentation of the results of the application of the method to different systems. The last section concludes the presented method.

\section{Class identification procedure}
The idea of class identification is independent on the type of the network we deal with. If only the analysed system is characterised by some kind of symmetry, which is the case for many real systems, it is possible to indicate groups of nodes which have the same properties in the network. Here, we are interested in properties which manifest as the same patterns of connections of particular nodes. As it was already mentioned in the Introduction, the same pattern of connections of some group of nodes means that they have the same number of nearest neighbours which are of the same kind. Further, if the considered network is weighted, not only the number of connections must be taken into account but also the weights of particular connections. The exact procedure differs slightly depending on the type of graph, as one has to take into account whether considered graph is directed and/or weighted. It should be also emphasized that even though we take explicitly into account only nearest neighbours of each node, in 
fact the procedure preserves information about connections with further neighbours. This is because the class of any node depends on the classes of its neighbours, which in turn depend on the classes of their neighbours and so on. A similar concept, termed 'regular equivalence', has been described in \cite{bor}.

\subsection{Simple graphs}
The simplest case is a graph which is undirected and unweighted. An example network is presented in Tab.\ref{tab_uu}. In the first column of the table, the node index appears, while in the next column there are indices of nearest neighbours of this node. As considered network is undirected and unweighted, the table provides full information needed for the class identification. The subsequent steps of the procedure are presented in Tab.\ref{cl_uu}. At the very beginning one has to assign the same symbol to all nodes which have the same degree (i.e. the same number of nearest neighbours). So, the number of different symbols equals the number of different values of the node degree, present in the considered network. In our example this number is equal to $5$, and we use letters from $A$ to $E$ to indicate nodes with different degrees. Now, also the numbers which identify neighbours of nodes should we replaced by just assigned symbols. This is done in the first part of Table \ref{cl_uu}. 
At the moment, one has to check whether symbols assigned to the neighbours of nodes which were decorated by given symbols are the same (it should be emphasized that not all neighbours of a given node should belong to the same class). If it is the case, this means that proper classes are already obtained. In the other case, the procedure is to be continued, as one step is not enough. In our example, in the case of nodes which have degree $4$ and were decorated by the symbol $B$, in one case neighbours are decorated by the symbol $D$ and in another case by the symbol $E$, as in the second and the last line of the first part of Table \ref{cl_uu}. This means that however degrees of those two nodes are the same, the patterns of their connections with other nodes of the network are different. One has to introduce further distinction of those nodes. Here we add a number which follows a given class symbol. As only nodes decorated by the symbol $B$ have to be distinguished, next to all remaining symbols we have $1$, and after $B$ we have $1$ and $2$. The introduced change involves an update of all already assigned symbols, which is done and presented in the second part of Tab.\ref{cl_uu}. The last step ensures the proper class identification for the analysed network. As the result, presented in Tab.\ref{clf_uu}, we obtain that $16$ possible states are classified to $6$ classes.

\begin{table*}
\begin{tabular}{c|*9{c}}
node number&\multicolumn{9}{c}{neighbours of a given node}\\\hline
0&2&4&5&6&7&9&&&\\
1&2&4&5&7&&&&&\\
2&0&1&3&9&10&11&12&&\\
3&2&4&5&6&7&9&&&\\
4&0&1&3&6&10&11&12&&\\
5&0&1&3&6&10&11&14&&\\
6&0&3&4&5&8&12&14&15&\\
7&0&1&3&9&10&11&14&&\\
8&6&9&&&&&&&\\
9&0&2&3&7&8&12&14&15&\\
10&2&4&5&7&12&14&&&\\
11&2&4&5&7&12&14&&&\\
12&2&4&6&9&10&11&13&15&\\
13&12&14&&&&&&&\\
14&5&6&7&9&10&11&13&15&\\
15&6&9&12&14&&&&&\\
\end{tabular}
\caption{The example of the undirected and unweighted graph}
\label{tab_uu}
\end{table*}

\begin{table*}
\begin{tabular}{c|*8{c}}
node class&\multicolumn{8}{c}{classes of neighbouring nodes}\\\hline
\multicolumn{9}{c}{first iteration}\\\hline
C&D&D&D&D&E&E\\
B&D&D&D&D\\
D&B&C&C&C&C&E&E\\
C&D&D&D&D&E&E\\
D&B&C&C&C&C&E&E\\
D&B&C&C&C&C&E&E\\
E&A&B&C&C&D&D&E&E\\
D&B&C&C&C&C&E&E\\
A&E&E\\
E&A&B&C&C&D&D&E&E\\
C&D&D&D&D&E&E\\
C&D&D&D&D&E&E\\
E&A&B&C&C&D&D&E&E\\
A&E&E\\
E&A&B&C&C&D&D&E&E\\
B&E&E&E&E\\\hline
\multicolumn{9}{c}{second iteration}\\\hline
C1&D1&D1&D1&D1&E1&E1\\
B1&D1&D1&D1&D1\\
D1&B1&C1&C1&C1&C1&E1&E1\\
C1&D1&D1&D1&D1&E1&E1\\
D1&B1&C1&C1&C1&C1&E1&E1\\
D1&B1&C1&C1&C1&C1&E1&E1\\
E1&A1&B2&C1&C1&D1&D1&E1&E1\\
D1&B1&C1&C1&C1&C1&E1&E1\\
A1&E1&E1\\
E1&A1&B2&C1&C1&D1&D1&E1&E1\\
C1&D1&D1&D1&D1&E1&E1\\
C1&D1&D1&D1&D1&E1&E1\\
E1&A1&B2&C1&C1&D1&D1&E1&E1\\
A1&E1&E1\\
E1&A1&B2&C1&C1&D1&D1&E1&E1\\
B2&E1&E1&E1&E1\\
\end{tabular}
\caption{Steps of classes identification procedure for the example of the undirected and unweighted graph presented in Tab.\ref{tab_uu}}
\label{cl_uu}
\end{table*}

\begin{table*}
\begin{tabular}{c|*8{c}}
node class&\multicolumn{8}{c}{classes of neighbouring nodes}\\\hline
A1&E1&E1\\
B1&D1&D1&D1&D1\\
B2&E1&E1&E1&E1\\
C1&D1&D1&D1&D1&E1&E1\\
D1&B1&C1&C1&C1&C1&E1&E1\\
E1&A1&B2&C1&C1&D1&D1&E1&E1\\
\end{tabular}
\caption{The final classes set for the example of the undirected and unweighted graph presented in Tab.\ref{tab_uu}}
\label{clf_uu}
\end{table*}

\subsection{Weighted graphs}
Consider the same network as in the previous subsection but with different weights of the edges. The modified network is presented in Tab.\ref{tab_uw}. For simplicity we consider just two different weight values, marked in the table by two different font styles. Now, not only the number of connections between nodes must be taken into account but also their weights. The procedure of classes identification also in this case starts with an assignment of symbols to the nodes to indicate their different degrees. As in the previous case, we have $5$ different values of the node degree. Both nodes decorated by the symbol $A$ have exactly the same kinds of neighbours and we do not need to introduce their further distinction. In the case of nodes decorated by the symbol $B$, we observe exactly the same situation as before; one of the nodes have neighbours which are decorated by the symbol $D$ and the other have neighbours which are decorated by the 
symbol $E$. This of course means that they are not the same, so we add to the symbol two different numbers to enable their distinction. We also see, that however the symbols assigned to the neighbours of nodes decorated by symbols $C$ and $D$ are the same, the weights of some edges are different. Because of that we also add additional number to both those symbols. $E$ nodes are the same so they do not need to be distinguished. After the second iteration of the procedure we see that the classes identification is not finished, because of the difference of classes, the neighbours of $E$ nodes belong to. The remaining symbols are at the moment correct, so we add the additional number $1$ to all of them, while one of $E$ nodes gets added $1$ and the another gets $2$. This is the third iteration in Tab.\ref{cl_uw}. The introduced changes are taken into account also in the lists of the neighbours of all nodes. Still, we did not obtain the proper classification. The last modification causes the necessity of a distinction of nodes decorated by the symbol $A$. After this modification we obtain the proper set of classes for the analysed network. The final result is presented in Tab.\ref{clf_uw}. In this case the original state space size equal to $16$ is reduced to $10$ classes.

\subsection{Directed graphs, unweighted and weighted}
\label{duw}
\begin{figure}[!hptb]
\begin{center}
\includegraphics[width=.5\columnwidth, angle=0]{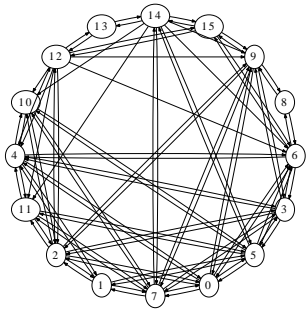}
\caption{The example of the directed and unweighted graph}
\label{gr_du}
\end{center}
\end{figure}

\begin{figure}[!hptb]
\begin{center}
\includegraphics[width=.5\columnwidth, angle=0]{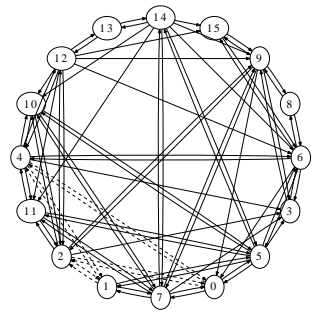}
\caption{The example of the directed and weighted graph, by dashed lines edges with different weights are marked}
\label{gr_dw}
\end{center}
\end{figure}

In the case of directed graphs we have to indicate not only lists of nodes which are obtained from a given node, but also lists of nodes which lead to a given node. The whole procedure is then the same as in the case described above; the only difference is that now two lists of neighbours are taken into account. Two examples, for unweighted and weighted graphs, are presented in Fig.\ref{gr_du} and Fig.\ref{gr_dw}, respectively. The final results of the class identification procedure are presented in Tab.\ref{clf_du} and Tab.\ref{clf_dw}. We see that in the case of directed and unweighted graph the system size is reduced to $9$ classes, and in the case of directed and weighted graph to $12$ classes.\\

\begin{table*}
\begin{tabular}{c|*9{c}}
node number&\multicolumn{9}{c}{neighbours of a given node}\\\hline
0&\textbf{2}&4&\textbf{5}&6&7&9&&&\\
1&2&4&5&7&&&&&\\
2&\textbf{0}&1&\textbf{3}&9&10&11&12&&\\
3&\textbf{2}&4&\textbf{5}&6&7&9&&&\\
4&0&1&3&6&10&11&12&&\\
5&\textbf{0}&1&\textbf{3}&6&10&11&14&&\\
6&0&3&4&5&8&12&14&15&\\
7&0&1&3&9&10&11&14&&\\
8&6&9&&&&&&&\\
9&0&2&3&7&8&12&14&15&\\
10&2&4&5&7&12&14&&&\\
11&2&4&5&7&12&14&&&\\
12&2&4&6&9&10&11&13&15&\\
13&12&14&&&&&&&\\
14&5&6&7&9&10&11&13&15&\\
15&6&9&12&14&&&&&\\
\end{tabular}
\caption{The example of the undirected and weighted graph; in bold connections with different weights are marked}
\label{tab_uw}
\end{table*}

\begin{table*}
{\scriptsize{
\begin{tabular}{ccc}
\hspace{-2cm}
\begin{tabular}{c|*8{c}}
node class&\multicolumn{8}{c}{classes of neighbouring nodes}\\\hline
\multicolumn{9}{c}{first iteration}\\\hline
C&D&D&\textbf{D}&\textbf{D}&E&E\\
B&D&D&D&D\\
D&B&C&C&\textbf{C}&\textbf{C}&E&E\\
C&D&D&\textbf{D}&\textbf{D}&E&E\\
D&B&C&C&C&C&E&E\\
D&B&C&C&\textbf{C}&\textbf{C}&E&E\\
E&A&B&C&C&D&D&E&E\\
D&B&C&C&C&C&E&E\\
A&E&E\\
E&A&B&C&C&D&D&E&E\\
C&D&D&D&D&E&E\\
C&D&D&D&D&E&E\\
E&A&B&C&C&D&D&E&E\\
A&E&E\\
E&A&B&C&C&D&D&E&E\\
B&E&E&E&E\\\hline
\multicolumn{9}{c}{third iteration}\\\hline
C21&D11&D11&\textbf{D21}&\textbf{D21}&E12&E12\\
B11&D11&D11&D21&D21\\
D21&B11&C11&C11&\textbf{C21}&\textbf{C21}&E11&E12\\
C21&D11&D11&\textbf{D21}&\textbf{D21}&E12&E12\\
D11&B11&C11&C11&C21&C21&E11&E12\\
D21&B11&C11&C11&\textbf{C21}&\textbf{C21}&E11&E12\\
E12&A11&B21&C21&C21&D11&D21&E11&E11\\
D11&B11&C11&C11&C21&C21&E11&E12\\
A11&E12&E12\\
E12&A11&B21&C21&C21&D11&D21&E11&E11\\
C11&D11&D11&D21&D21&E11&E11\\
C11&D11&D11&D21&D21&E11&E11\\
E11&A11&B21&C11&C11&D11&D21&E12&E12\\
A11&E11&E11\\
E11&A11&B21&C11&C11&D11&D21&E12&E12\\
B21&E11&E11&E12&E12\\
\end{tabular}
&&
\begin{tabular}{c|*8{c}}
node class&\multicolumn{8}{c}{classes of neighbouring nodes}\\\hline
\multicolumn{9}{c}{second iteration}\\\hline
C2&D1&D1&\textbf{D2}&\textbf{D2}&E1&E1\\
B1&D1&D1&D2&D2\\
D2&B1&C1&C1&\textbf{C2}&\textbf{C2}&E1&E1\\
C2&D1&D1&\textbf{D2}&\textbf{D2}&E1&E1\\
D1&B1&C1&C1&C2&C2&E1&E1\\
D2&B1&C1&C1&\textbf{C2}&\textbf{C2}&E1&E1\\
E1&A1&B2&C2&C2&D1&D2&E1&E1\\
D1&B1&C1&C1&C2&C2&E1&E1\\
A1&E1&E1\\
E1&A1&B2&C2&C2&D1&D2&E1&E1\\
C1&D1&D1&D2&D2&E1&E1\\
C1&D1&D1&D2&D2&E1&E1\\
E1&A1&B2&C1&C1&D1&D2&E1&E1\\
A1&E1&E1\\
E1&A1&B2&C1&C1&D1&D2&E1&E1\\
B2&E1&E1&E1&E1\\\hline
\multicolumn{9}{c}{fourth iteration}\\\hline
C211&D111&D111&\textbf{D211}&\textbf{D211}&E121&E121\\
B111&D111&D111&D211&D211\\
D211&B111&C111&C111&\textbf{C211}&\textbf{C211}&E111&E121\\
C211&D111&D111&\textbf{D211}&\textbf{D211}&E121&E121\\
D111&B111&C111&C111&C211&C211&E111&E121\\
D211&B111&C111&C111&\textbf{C211}&\textbf{C211}&E111&E121\\
E121&A112&B211&C211&C211&D111&D211&E111&E111\\
D111&B111&C111&C111&C211&C211&E111&E121\\
A112&E121&E121\\
E121&A112&B211&C211&C211&D111&D211&E111&E111\\
C111&D111&D111&D211&D211&E111&E111\\
C111&D111&D111&D211&D211&E111&E111\\
E111&A111&B211&C111&C111&D111&D211&E121&E121\\
A111&E111&E111\\
E111&A111&B211&C111&C111&D111&D211&E121&E121\\
B211&E111&E111&E121&E121\\
\end{tabular}
\end{tabular}
\caption{Steps of classes identification procedure for the example of the undirected and weighted graph presented in Tab.\ref{tab_uw}; in bold connections with different weights are marked}
\label{cl_uw}
}}
\end{table*}

\begin{table*}
\begin{tabular}{c|*8{c}}
node class&\multicolumn{8}{c}{classes of neighbouring nodes}\\\hline
A111&E111&E111\\
A112&E121&E121\\
B111&D111&D111&D211&D211\\
B211&E111&E111&E121&E121\\
C111&D111&D111&D211&D211&E111&E111\\
C211&D111&D111&\textbf{D211}&\textbf{D211}&E121&E121\\
D111&B111&C111&C111&C211&C211&E111&E121\\
D211&B111&C111&C111&\textbf{C211}&\textbf{C211}&E111&E121\\
E111&A111&B211&C111&C111&D111&D211&E121&E121\\
E121&A112&B211&C211&C211&D111&D211&E111&E111\\
\end{tabular}
\caption{The final class set for the example of the undirected and weighted graph presented in Tab.\ref{tab_uw}; in bold connections with different weights are marked}
\label{clf_uw}
\end{table*}

\begin{table*}
\begin{tabular}{c|*8{c}|*8{c}}
node&\multicolumn{16}{c}{classes of neighbouring nodes}\\
class&\multicolumn{8}{c|}{in-neighbours}&\multicolumn{8}{c}{out-neighbours}\\\hline
A1&E1&E1& & & & & & &E1&E1\\
A2&G1&G1& & & & & & &G1&G1\\
B1&E1&E1&G1&G1& & & & &E1&E1&G1&G1\\
B2&F1&F1&F1&F1& & & & &F1&F1&F1&F1\\
C1&E1&E1&F1&F1&F1&F1& & &F1&F1&F1&F1\\
D1&F1&F1&F1&F1&G1&G1& & &F1&F1&F1&F1&G1&G1\\
E1&A1&B1&F1&F1&G1&G1& & &A1&B1&C1&C1&F1&F1&G1&G1\\
F1&B2&C1&C1&D1&D1&E1&G1& &B2&C1&C1&D1&D1&E1&G1\\
G1&A2&B1&D1&D1&E1&E1&F1&F1&A2&B1&D1&D1&E1&E1&F1&F1\\
\end{tabular}
\caption{The final class set for the example of the directed and unweighted graph presented in Fig.\ref{gr_du}}
\label{clf_du}
\end{table*}

\begin{table*}
{\scriptsize{
\begin{tabular}{c|*7{c}|*8{c}}
node&\multicolumn{15}{c}{classes of neighbouring nodes}\\
class&\multicolumn{7}{c|}{in-neighbours}&\multicolumn{8}{c}{out-neighbours}\\\hline
A11 &B11 &B12     &       &        &       &  &  &B11 &B12\\
A21 &H11 &H11     &       &        &       &  &  &H11 &H11\\
B11 &A11 &G11     &G11     &        &       &  &  &A11 &C11 &F11 &F11 &G11 &G11 &H11 &H11\\
B12 &A11 &G21     &G21     &        &       &  &  &A11 &C11 &F11 &F11 &G21 &G21 &H11 &H11\\
C11 &B11 &B12     &H11     &H11      &       &  &  &H11 &H11\\ 
D11 &G11 &G11     &\textbf{G21} &\textbf{G21}  &       &  &  &G11 &G11 &\textbf{G21} &\textbf{G21}\\
E11 &G11 &G11     &G21     &G21      &H11     &H11 &  &H11 &H11\\
F11 &B11 &B12     &G11     &G11      &G21     &G21 &  &G11 &G11 &G21 &G21\\
F21 &G11 &G11     &\textbf{G21} &\textbf{G21}  &H11     &H11 &  &G11 &G11 &\textbf{G21} &\textbf{G21}\\
G11 &B11 &D11     &F11     &F11      &F21     &H11 &  &B11 &D11 &E11 &F11 &F11 &F21 &H11\\
G21 &B12 &\textbf{D11} &F11     &F11      &\textbf{F21} &H11 &  &B12 &\textbf{D11} &E11 &F11 &F11 &\textbf{F21} &H11\\
H11 &A21 &B11     &B12     &C11      &E11     &G11 &G21 &A21 &C11 &E11 &F21 &G11 &G21\\
\end{tabular}
\caption{The final class set for the example of the directed and weighted graph presented in Fig.\ref{gr_dw}, in bold connections with different weights are marked}
\label{clf_dw}
}}
\end{table*}

The rate of the system size reduction depends on the structure of the particular network. However, as it will be presented in Sec.\ref{res} in the case of large, real systems it can be significant.

\section{Equivalence of the representation of the system by classes to the representation by states}
The usefulness of the method arises from the fact that once we are interested in properties of the system in the stationary state, the representation of the system by means of classes is equivalent to the representation of the system by means of states. The core of this equivalence is that the probability of a given class equals to the sum of the probabilities of states which form this class. In the simplest case, if all states of the system are equally probable, the class probability $p_c$ is expressed as $p_c=n/N$, where $n$ is the number of states which belong to a given class and $N$ is the system size: the number of nodes. In general, to determine the probabilities of particular states one can calculate them from the transition matrix \cite{kamp}. This matrix should be constructed both for the network of states and the network of classes. In the transition matrix, an $i-th$ column is related to the rates of possible transitions from an $i-th$ state/class to other states/classes of the system. The probabilities of particular states/classes are obtained from the components of the normalized dominant eigenvector connected with the dominant eigenvalue $1$ of the appropriate transition matrix. As a criterion of the correctness of the obtained classification of the state space into classes, the equality may be checked of the probabilities of all states which belong to a given class, and the probabilities of particular class; the latter is equal to the probability of the states it contains multiplied by the number of states it contains.\\

This method of the probability determination cannot be however applied, if more than one absorbing state are present in the system; these states act as traps. An absorbing state is a state which can be obtained from other states of the system but itself cannot be transformed to any other state. In such a case, the transition matrix contains columns which consist of zeros except one unity as the diagonal term. In this case, the final stationary probability distribution depends on the initial distribution, and the transition matrix cannot be the base of the calculation of the probabilities. To solve the problem the set of Master equations \cite{kamp} can be used: \[\dfrac {dP_i (t)}{dt}=\sum\limits_{j\in S_i}P_j (t)w_{j\rightarrow i}- \sum\limits_{j\in S_i}P_i (t)w_{i\rightarrow j}\] In the above equation we sum over the set of nearest neighbours of a given node $i$, and $w$ is a weight of a given edge in a considered graph. Then, the probabilities can be found from its solution in the asymptotically long time, given the initial distribution.\\

For some directed networks it is possible that some state are neither end points of the evolution nor elements of limit cycles. Accordingly, the probabilities of these states tend to zero in time. One of our examples are cellular automata where these states are known as "Garden of Eden" \cite{wolf}.

\section{Examples of applications of the method}
\label{res}
As it was mentioned above, the method of the reduction of the system size by its representation by means of classes is general. Below we present a brief review of systems where the method has been applied. We also present main results obtained for each system. Details are presented in the cited literature. 
\begin{enumerate}
\item\textit{Ising and Potts models in spatial structures with geometrical frustration} \cite{mk1} We analyse all periodic states of small pieces of some spatial structures for the Ising model with two possible spin orientations and the Potts model with three possible spin orientations, with antiferromagnetic interaction. The first analysed system is the triangular lattice with the periodic boundary conditions. In this case each node has six neighbours. The system properties, including the frustration effect, strongly depend on the size of the system and the model used. The second analysed lattice is the cubic Laves phase $C15$. An example is the $YMn_2$ intermetallic \cite{ymn}, where $16$ Mn atoms form four tetrahedra. Similarly to the triangular lattice, each node has six neighbours, but their mutual positions form a three-dimensional structure. The structure of the lattice causes spins in the system to be frustrated, both for $2$ and $3$ possible spin orientations. For both systems, all ground states are listed which fulfil the periodic boundary conditions. Those ground states are used as nodes of the constructed network, while edges are defined by single spin flips which transform one ground state of the system to another. The obtained network is undirected and unweighted. The reduction of the system size due to symmetry is possible in the case of the triangular lattice with Ising antiferromagnetic interactions; there, the system size is reduced from $3\;630$ states to $12$ classes for $25$ atoms, and from $263\;640$ states to $409$ classes for $36$ atoms. In the case of the Potts model in $C15$ structure, our procedure allows us to reduce the system size from $90\;936$ states to $28$ classes.

\item\textit{Traffic system} \cite{mk2} We analyse the network formed with the state space of vehicles on a small roundabout. In our system, there are three access and three exit roads. The maximal number of vehicles on each road is equal to $2$, so on each road two, one, or no cars is permitted. The state of the system is denoted as a sequence of six signs, where each sign reflects the current state of one road. As at most $2$ vehicles are allowed on each road at the same time, a new car can appear on an access road only if its current occupation is less than two. Neither the change of a state of any access road from $0$ to $1$ or from $1$ to $2$ nor the decrease of the number of vehicles on an exit road do not influence the state of remaining roads of the system. A shift of a vehicle from an access to an exit road is possible, if an access road is occupied at least by one car and an exit road is occupied by at most one car. When a roundabout is empty, a car can appear on any of access roads, with equal 
probability. If at least one vehicle is on an access road, a passage is possible through the roundabout to each of the exit roads occupied by at most one vehicle. The probability of a passage to a given exit road depends on the distance to cover (the distance on the roundabout passed by the car is treated by analogy to the electric resistance). Assumed values of the model parameters - the number of roads and the number of permitted vehicles - cause that the whole state space contains $3^6=729$ states which determine nodes of the network. Any two nodes (states) are connected if a change of the position of one vehicle transforms one of those states into another. The obtained network of states is directed and weighted. We note here that, unfortunately, the results presented in our paper \cite{mk2} are inconsistent with what was written there in the model description. In fact, the division of 729 possible states of the system into $55$ classes, reported in \cite{mk2}, occurs for the case when weights of all passages are equal. If it is not the case, and passages from 
input to output roads have different weights - as described above - $138$ classes are obtained.

\item\textit{Polymer chain} \cite{mk4} We analyse the state space formed by the set of all possible conformations of a circular polymer molecule in the repton model \cite{deg,new,el1,el2}. We assume that reptons are indistinguishable, so in a consequence a shift of the whole molecule along its contour does not lead to a different state. What we however distinguish are the lattice cells. Yet, a translation of the whole molecule in the lattice does not generate a new state. With these assumptions, we find all possible conformations of a molecule of a given length. The length of the molecule is expressed by the number of reptons $N$. An obvious symmetry of the system, which allows  for the reduction of the system size, is the symmetry of reflection in the lattice axes; all states of the reflected shape are equivalent to their initial counterparts. However, our method allows us to take into account also less trivial symmetries resulting from the similarities in the properties of particular states which are connected with possible transitions between the states. These similarities concern both the number of possible transitions and their probabilities. As the size of the state space depends on the molecule length and the probabilities of transitions depend on the strength and direction of the external electric field, our method allows for an analysis of the system in different motion regimes: strong field, weak field and zero field. Yet, in each case the size of the system can be reduced. The reduction rate depends on the current values of the model parameters. In the case  of zero electric field the network of states is undirected and unweighted. For a molecule consisted of $N=5$ reptons, $35$ states are reduced to $9$ classes, and for $N=9$, $4891$ states are reduced to $702$ classes. If the molecule is placed in a gel medium, for $N=5$ we get $31$ molecule conformations; the number of classes is equal to $6$. In the case with field, in which case the network of states is directed and weighted, the number of classes depends on the direction of the applied field with respect to the square lattice. If the direction of the field is parallel to the lattice axes, $21$ classes for the $5$-repton long molecule are identified. A slightly smaller number of classes - $14$ - is identified for field directions parallel to the diagonal of the lattice cells.

\item\textit{Elementary cellular automata} \cite{mk5} We analyse one-dimensional cellular automata, with two possible states of each cell. The rule of the change of a cell state is determined on the basis of the state of the cell itself and its two nearest neighbours. Such a definition leads to $2^8=256$ different rules of the evolution process, so called elementary cellular automata \cite{wolf1}. It has been shown that the number of unique rules is lower, as some of the rules are equivalent. As a result, the set of all elementary automata is represented by $88$ unique rules \cite{wolf1}. We consider a system of a specified length $N$ consisting of $2^N$ possible sequences of zeros and ones, with periodic boundary conditions. Each sequence is converted to a single sequence determined by the automaton rule. Each sequence may, however, be obtained by the transformation of different sequences. The exact number of states (sequences) leading to a given state depends on the automaton rule and the system size $N$. 
The state space can be represented as a directed unweighted graph, where $2^N$ nodes are identified with the possible states of the system. The out-degree of all nodes is equal to $1$, as each state can be transformed to only one other state, while the in-degree is different for different nodes. We have shown in \cite{mk5} that different character of the relation of the number of classes $\#$ to the system size $N$ are observed. For most of the rules, an exponential increase is observed, but we also observe some number of rules which show different kinds of behaviour. Namely, besides the obvious case of automata No $0$ and $255$, a non-trivial dependence of the number of classes in a function of the system size is observed for automata No $15, 45, 60, 90, 105$ and $ 154$ (and their equivalent rules). In most cases, the class identification allows a significant reduction of the system size. The proposed method indicates that the number of symmetry groups is equal to $80$, as for some groups indicated as different 
by Wolfram, the same pattern $\#(N)$ is observed.

\item\textit{Hubbard ring} \cite{mk6} We analyse the Hubbard ring, which is an example of a quantum system. We consider a ring, i.e. a one-dimensional circular chain of atoms, within the single-band Hubbard model in the atomic limit \cite{ham}. In accordance with the Pauli principle, each atom can be occupied by at most two electrons, and if it is the case their spins are opposite. Then, for a given chain length we can find all possible states; they form the state space of the analysed system. The set of states can be represented in the form of a network, where each state of the system can be treated as a node. An edge appears between two nodes if one state can be obtained from another in one of two possible processes: electron hopping and spin flip. The rates of those two processes are different, and the rate of the hopping is expected to be higher than the rate of flips \cite{kha}. Further, a spin can flip only if at most a single electron is located at the target atom, also because of the Pauli principle. 
Here we take into account only hopping between neighbouring atoms of the ring, as further hoppings have very low probabilities \cite{ham}. The electron hopping occurs with no change of the spin orientation. The obtained network is undirected and weighted. In our approach, the translational symmetry is taken into account explicitly, i.e. we do not distinguish the states which can be obtained one from another by a shift along the ring.\\
In our paper we analysed four cases: the full set of states, the ground states, and in both cases the state space with or without duplicates which are mirror reflections of other states of the system. In all cases the number of classes to the number of states ratio depends on the occupation ratio (number of electrons to doubled number of atoms). Also, which is intuitively clear, the difference is seen for cases with and without mirror reflections. However, in all cases the reduction ratio is significant, namely from $0.75$ to $0.003$.\\
\end{enumerate}

\section{Boolean networks}
The list of examples brought up in the preceding section provides a summary of the up-to-date applications of the method. Below we add a new application to the random Boolean network  \cite{dros,kauf}. These networks are of interest for their relevance for genetic networks \cite{born}.\\

The state space consists of a set of all possible states of a Boolean network of a given size $N$, where each node can be in one of two possible states, let us say $0$ and $1$. This means that the size of the obtained state space equals $2^N$. The definition of the Boolean network requires an indication of the neighbours of each node, which is made by the random assignment of the set of $k$ nodes to each node, while a node cannot be a neighbour of itself. The state of each node changes in accordance to a Boolean function, which indicates the state of a given node in the next time step, based on the states of its neighbours. As each node has $k$ neighbours, there are $2^{2^k}$ possible Boolean functions, one of which is randomly assigned to each node. We assume that these functions do not change over time, and the states of all nodes are updated in parallel. The change of a state of all nodes leads to another state of the state space. As the process is deterministic, each state is transformed to exactly one state, while given state can be obtained from more states. The obtained graph is directed and unweighted. In accordance with what was written in Sec.\ref{duw}, the class identification procedure must take into account both in- and out-neighbours lists.\\

The Boolean networks are somehow similar to the cellular automata analysed in our other paper \cite{mk5}, however in the latter case a state of the node in the next generation depends not only on the states of its neighbours but also on the state of itself. The main difference is that the rules of cellular automata are the same for each cell, while in the case of the Boolean networks the change of the node state depends on the Boolean function chosen for this node, so for the same structure of the network the connections between particular states may be different. \\

\begin{figure}[!hptb]
\begin{center}
\includegraphics[width=.8\columnwidth, angle=270]{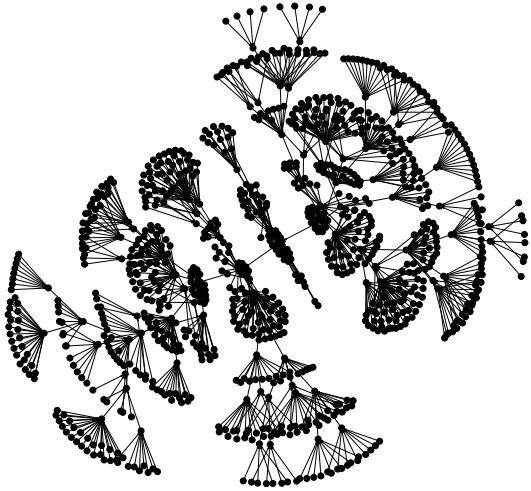}
\caption{Example network for the state space of the size $N=2^{10}$ states, formed from the Boolean networks for $n=10$ and $k=3$, where $n$ - number of nodes, $k$ - node degree.}
\label{netS}
\end{center}
\end{figure}

\begin{figure}[!hptb]
\begin{center}
\includegraphics[width=.7\columnwidth, angle=270]{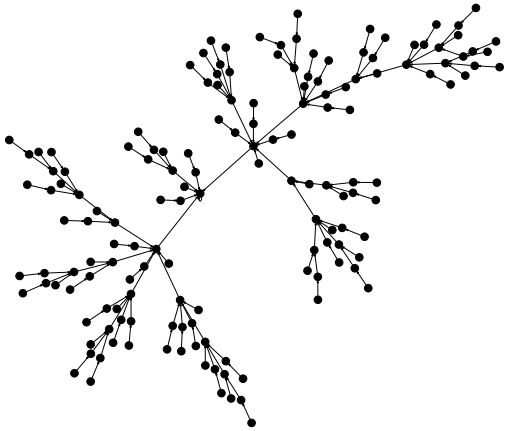}
\caption{Network of classes for the network presented in Fig.\ref{netS}.}
\label{netC}
\end{center}
\end{figure}

An example of the network for the state space of the size $N=2^{10}$ states, formed from the Boolean networks for $n=10$ and $k=3$, where $n$ is the number of nodes and $k$ is the node degree is presented in Fig.\ref{netS}. The related network of classes is presented in Fig.\ref{netC}.

\begin{figure}[!hptb]
\begin{center}
\includegraphics[width=.8\columnwidth, angle=270]{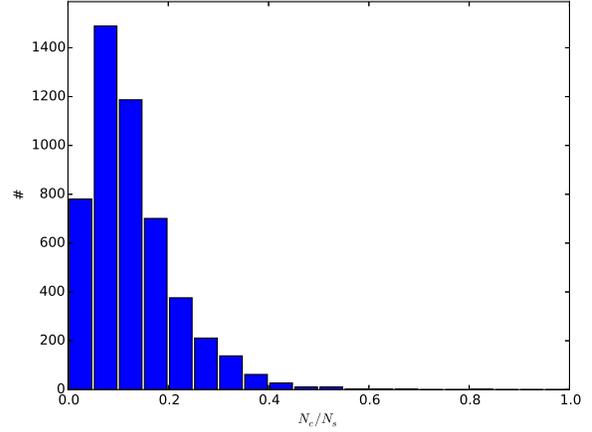}
\caption{Reduction of the system size $N_c/N_s$ for the random Boolean networks for $N=2^{10}$ states and $k=2$ for $5\times10^3$ repetitions, where $k$ - node degree, $N_c$ - number of classes and $N_s$ - number of states.}
\label{N10_k2}
\end{center}
\end{figure}

\begin{figure}[!hptb]
\begin{center}
\includegraphics[width=.8\columnwidth, angle=270]{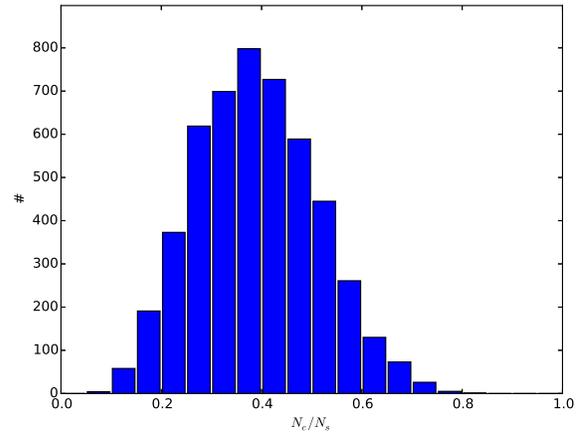}
\caption{Reduction of the system size $N_c/N_s$ for the random Boolean networks for $N=2^{10}$ states and $k=3$ for $5\times10^3$ repetitions, where $k$ - node degree, $N_c$ - number of classes and $N_s$ - number of states.}
\label{N10_k3}
\end{center}
\end{figure}

Fig.\ref{N10_k2} presents the reduction of the system size expressed as the ratio $N_c/N_s$, where $N_c$ number of classes and $N_s$ number of states, for the space state of the Boolean networks which consists of $N=2^{10}$ states with node degree $k=2$, and for $5\times10^3$ repetitions. The average value is equal to $0.13\pm0.09$. If we change the node degree in the Boolean networks to $k=3$, the reduction rate is lower, as presented in Fig.\ref{N10_k3}. In this case, $<N_c/N_s>=0.39\pm0.12$. In the analysed set of $5\times10^3$ networks in $90$ percent of them cycles are observed.

\section{Discussion}
The goal of the presented method is that it allows us to reconstruct the probability distribution of the states of the analysed system. This is done as follows: at first the graph which reflects the mutual relations between the elements of the system must be constructed. Then, the obtained graph is used for the indication of classes of states, which causes the reduction of the graph, and the equivalent representation of the system by classes. For the obtained network of classes one can calculate the probability distribution. When this is done, the probability distribution of the original system can be obtained, as the class probability is equal to the probability of states which form a given class multiplied by the number of states in this class.\\
As it was shown the presented method of the symmetry-driven reduction of the system size is general, and it can be applied to different kinds of discrete systems, not only being the subject of the classical physics. Here, we have presented six examples of systems, where the method has been applied. In all cases the method ensures that the probability of each state which belongs to a given class is the same. Thus, any analysis which concerns a stationary probability distribution of states of the analysed system can be performed using its reduced, equivalent, representation. The rate of the system size reduction depends on the considered system but in most cases it is significant.

\section*{Acknowledgments}
The author is grateful to Krzysztof~Kułakowski for critical reading of the manuscript and helpful discussions. The author is also grateful to the anonymous Referee for valuable comments that improved the manuscript. This research was supported in part by PL-Grid Infrastructure and in part by the Ministry of Science and Higher Education (MNiSW).

\end{document}